\documentclass[aps,pra,showpacs,twocolumn]{revtex4-1}
\usepackage{amsmath}
\usepackage{amssymb}
\usepackage{graphicx}
\usepackage{hyperref}
\newcommand{\ket}[1]{|#1\rangle}
\newcommand{\bra}[1]{\langle#1|}

\begin{document}
%\title{Existence of a Kochen-Specker inequality is independent of the Kochen-Specker rules}
\title{Coexistence of Kochen-Specker inequalities and noncontextuality inequalities}
\author{Xiao-Dong Yu}
\affiliation{Department of Physics, Shandong University, Jinan 250100, China}
\author{D. M. Tong}
\email{tdm@sdu.edu.cn}
\affiliation{Department of Physics, Shandong University, Jinan 250100, China}
\date{\today}
\begin{abstract}
Two types of inequalities, Kochen-Specker inequalities and noncontextuality inequalities, are both used to demonstrate the incompatibility between the noncontextual hidden variable model and quantum mechanics. It has been thought that noncontextuality inequalities are much more potent than Kochen-Specker inequalities, since the latter are constrained by the Kochen-Specker rules, which are regarded as an extra constraint imposed on the noncontextual hidden variable model. However, we find that a noncontextuality inequality exists in a ray set if and only if a Kochen-Specker inequality exists in the same ray set.
%This finding indicates that both  Kochen-Specker and noncontextuality inequalities are equally effective in demonstrating the incompatibility.
This provides an effect approach both for constructing
noncontextuality inequalities in a Kochen-Specker set and for
converting a Kochen-Specker inequality to a noncontextuality
inequality in any ray set.
\end{abstract}
\pacs{03.65.Ta, 03.65.Ud} \maketitle

The noncontextual hidden variable (NCHV) model
\cite{Kochen.Specker1967}, as with the local hidden variable (LHV)
model \cite{Bell1966}, has attracted a lot of attention. The NCHV
model consists of two basic assumptions: that every observable $A$
has a definite value $v(A)$ at all times and that $v(A)$ does not
depend on whether $A$ is measured alone or together with $B$ or $C$
if $A$ is compatible with $B$ and $C$. The first assumption is at
the core of the hidden variable theory while the latter is the 
exhibition of noncontextuality. The most common observables used in the
NCHV model are rays, $P_i,~i=1,2,\cdots$. The NCHV model indicates
that $v(P_i)$ can take only $0$ or $1$.

To examine the inconsistency of the NCHV model with quantum mechanics,
the pioneering works
\cite{Kochen.Specker1967,Peres1991,Bub1996,Conway.Kochen,Kernaghan1994,Cabello.etal1996,Clifton1993}
used to impose the constraint of preserving the algebraic structure
of compatible observables, i.e., the so-called product rule
$v(AB)=v(A)v(B)$ and sum rule $v(A+B)=v(A)+v(B)$ for two compatible
observables $A$ and $B$. For rays, the constraint is equivalent to
the  Kochen-Specker (KS) rules: $v(P_i)v(P_j)=0$ if $P_i$ and $P_j$
are orthogonal, and exactly one $v(P_i)$ is $1$ out of an orthogonal
basis set of rays. A ray set $\mathcal{S}$ is called a KS set if all
possible value assignments to the ray set violate the KS
rules. By employing the KS rules, Kochen and Specker
\cite{Kochen.Specker1967} illustrated logically that the NCHV model
is inconsistent with quantum mechanics, which is usually called 
the KS theorem. Their original argument involved a KS set of $117$
rays in a three-dimensional Hilbert space. The proof of the KS theorem
is simplified step by step
\cite{Peres1991,Bub1996,Conway.Kochen,Kernaghan1994,Cabello.etal1996},
with smaller KS sets, e.g., with $31$ rays in a three-dimensional
space by Conway and Kochen \cite{Conway.Kochen} or with $18$ rays in
a four-dimensional space by Cabello \textit{et al.}
\cite{Cabello.etal1996}. To simplify further the logical proof, KS
inequalities, as a reformulation of the KS theorem, were then
proposed. Hereafter, when we use the term KS inequalities, we refer to those
inequalities in which the value assignments are constrained both by
the two basic assumptions in the NCHV model and by the KS rules \cite{Bengtsson.etal2012}. The
statistical approach proposed by Clifton \cite{Clifton1993} may be
regarded as the embryonic form of a KS inequality. Klyachko
\textit{et al.} \cite{Klyachko.etal2008} constructed a KS inequality
by using a set of only five rays in a three-dimensional space. More KS
inequalities can be found in recent papers
\cite{Simon.etal2001,Larsson2002,Cabello.etal2008,Yu.Oh2012,Bengtsson.etal2012,Cabello.etal2010,Kurzynski.Kaszlikowski2012}.

Since KS inequalities are not only based on the assumption of noncontextuality, but are also dependent on the KS rules, which is an extra assumption imposed on the NCHV model and difficult to test in experiment, researchers prefer to find inequalities without employing the KS rules. Such general inequalities, in which the value assignments are constrained only by the two basic assumptions in the NCHV model but not by the KS rules, are
called noncontextuality inequalities in order to differentiate them from KS inequalities \cite{Bengtsson.etal2012}. Klyachko
\textit{et al.} \cite{Klyachko.etal2008} found one such inequality, the KCBS inequality. Cabello
\cite{Cabello2008} constructed several state-independent noncontextuality inequalities, which are violated by all quantum
states. Soon afterwards, Badzi\c{a}g \textit{et al.} \cite{Badziag.etal2009} put forward a general approach to construct
a state-independent noncontextuality inequality from a KS set, which shows that any KS set can always lead to a noncontextuality inequality. Yu and Oh \cite{Yu.Oh2012} found a $13$-ray state-independent
noncontextuality inequality in a non-KS set. Some of these theoretical results have been demonstrated in experiments
\cite{Michler.etal2000,Huang.etal2003,Hasegawa.etal2006,Bartosik.etal2009,Liu.etal2009,Kirchmair.etal2009,Amselem.etal2009,Moussa.etal2010,Lapkiewicz.etal2011,Zu.etal2012,Zhang.etal2013}.

As discussed above, both KS inequalities and noncontextuality
inequalities have been used to demonstrate the incompatibility
between the NCHV model and quantum mechanics. It has been usually
thought that noncontextuality inequalities are much more potent than
KS inequalities, since the KS rules are regarded as an extra
constraint imposed on the NCHV model. However, it is much easier to
construct a KS inequality than a noncontextuality inequality in a
ray set, since a great number of the value assignments to the set
can be removed by the KS rules. Interestingly, we notice that in some
cases such as those in Refs.
\cite{Klyachko.etal2008,Yu.Oh2012,Bengtsson.etal2012}, both a KS
inequality and a noncontextuality inequality exist in the same ray
set. The result in Ref. \cite{Badziag.etal2009} also implies that
noncontextuality inequalities and KS inequalities coexist in a KS
set. This started us wondering whether a noncontextuality inequality
always coexists with a KS inequality in any ray set, not only in
some particular ray sets, and whether the former can be derived by
the aid of the latter if the coexistence is true. In this Rapid Communication, we
address these issues. We will prove the following theorem.

\textbf{Theorem}: A noncontextuality inequality exists in a ray set if and only if a Kochen-Specker inequality exists in the same ray set.

The theorem uncovers an essential relationship between KS
inequalities and noncontextuality inequalities. It clarifies that a
noncontextuality inequality always coexists with a KS inequality in
any ray set. The proof of theorem provides an effective approach
both for constructing a noncontextuality inequality in a KS set and
for converting a KS inequality to a noncontextuality inequality in
any ray set.

We now prove the theorem. To this end, we will first show that a noncontextuality inequality can be derived from a given KS inequality.

Consider an $n$-dimensional quantum system. Let $\mathcal{S}$ be a $\mu$-ray set of the system, $\mathcal{S}=\{P_1, P_2,\cdots,P_\mu\}$. $P_i$ have definite values $0$ or $1$ in the NCHV model, while they should be viewed as rank-1 projective operators $\hat{P_i}$ in quantum mechanics. Suppose that a KS inequality has been given as
 \begin{equation}
 \langle F\rangle_{ks} \le f,
 \label{eqF}
 \end{equation}
where $F\equiv F(P_1,P_2,\cdots,P_\mu)$ is a function of the observables $P_1, P_2,\cdots,P_\mu$, $\langle F\rangle_{ks}$ represents the average value of the function in the NCHV model, and $f$ is a constant number. Hereafter, the subscript $ks$ in $\langle \cdots\rangle_{ks}$ means that $\langle \cdots\rangle_{ks}$ is calculated under the constraint of the KS rules, i.e., the value assignments to the ray set $\{P_1,P_2,\dots,P_\mu\}$ obey the KS rules.

With the above knowledge, we may now start to derive a noncontextuality inequality from inequality (\ref{eqF}).

First, we introduce a new function, $\tilde{F}=\tilde{F}(P_1,P_2,\cdots,P_\mu$). The ray set $\mathcal{S}$ can be illustrated by a graph $\mathcal{G}$. Each vertex of $\mathcal{G}$ corresponds to a ray of $\mathcal{S}$, and two vertices are linked by an edge if and only if the corresponding rays are orthogonal. Besides, we use $\mathcal{S}^\alpha$, $\alpha=1,2,\dots,L$, to denote the $\alpha$th subset of the rays that can form an $n$-dimensional orthogonal basis, $\mathcal{S}^\alpha=\left\{ P_{k^\alpha_1},P_{k^\alpha_2},\dots,P_{k^\alpha_n} \right\}$, where $k^\alpha_1<k^\alpha_2<\cdots<k^\alpha_n$. $L$ is the total number of such subsets found in $\mathcal{S}$, and it is zero if $\mathcal{S}$ does not contain any $n$-dimensional orthogonal basis. The new function $\tilde{F}$ is defined as
\begin{equation}
\tilde{F}=\lambda F-\sum^\mu_{\substack{i>j=1,\\\hat{P_i}\hat{P_j}=0}}P_iP_j+
\sum_{\alpha=1}^L \left(\sum_{i=1}^n P_{k^\alpha_i}-2\sum_{i>j=1}^n P_{k^\alpha_i}P_{k^\alpha_j}\right),
  \label{edefBj}
\end{equation}
where $\lambda$ is a positive parameter to be determined. The first term in $\tilde{F}$ is the function $F$ defined in Eq. (\ref{eqF}), the second term corresponds to the pairs of rays that are orthogonal, and the third term corresponds to the sets of rays that form an $n$-dimensional orthogonal basis.

Second, we calculate the average value of $\tilde{F}$ under the NCHV model, with
\begin{eqnarray}
\langle \tilde{F}\rangle =&&\lambda\langle F\rangle-\sum^\mu_{\substack{i>j=1,\\\hat{P_i}\hat{P_j}=0}}\langle P_iP_j\rangle\nonumber\\
&&+\sum_{\alpha=1}^L \left(\sum_{i=1}^n \langle P_{k^\alpha_i}\rangle-2\sum_{i>j=1}^n\langle P_{k^\alpha_i}P_{k^\alpha_j}\rangle\right).
\label{average}
\end{eqnarray}
Clearly, $\langle \tilde{F}\rangle$ depends on the values of $P_i$ and $P_iP_j$. We first consider the case where the value assignments to the ray set obey the KS rules. Under the KS rules, $P_iP_j=0$ if $P_i$ and $P_j$ are orthogonal, and exactly one $P_i=1$ out of an orthogonal basis. We immediately have
\begin{eqnarray}
\tilde{F}|_{ks}=\lambda F|_{ks}+L,
\label{average2}
\end{eqnarray}
where $\tilde{F}|_{ks}$ ($F|_{ks}$) is the value of function $\tilde{F}$ ($F$). The subscript $ks$ means that $\tilde{F}|_{ks}$ ($F|_{ks}$) is obtained when the assignments of values $0,1$ to the ray set obey the KS rules. Noting that Eq. (\ref{eqF}) implies $F|_{ks}\le f$, we further have
\begin{eqnarray}
\tilde{F}|_{ks}\leq \lambda f+L.
\label{new2}
\end{eqnarray}
We now consider the case where the value assignments to the ray set violate the KS rules. The value of the expression $(\sum_{i=1}^n P_{k_i^\alpha}-2\sum_{i>j=1}^n P_{k^\alpha_i}P_{k^\alpha_j})$, denoted by $(\cdots)_\alpha$, is only dependent on the number of $1$'s (or $0$'s) assigned to the rays in the subset $\mathcal{S^\alpha}$ but independent of particular value assignments due to the symmetry in the expression. Thus $(\cdots)_\alpha=x(2-x)$, if the number of $1$'s in the value assignments is $x$. It shows that $(\cdots)_\alpha\leq 0$ if $x\ne 1$. Therefore, if the KS rules are violated, at least one $P_iP_j=1$ or one $(\cdots)_\alpha\le 0$. We then have
\begin{eqnarray}
\tilde{F}|_{exks}\leq \lambda F|_{exks}+L-1, \label{new3}
\end{eqnarray}
where the subscript $exks$ means that the values are obtained in the case where the value assignments to the ray set $\mathcal{S}$ violate the KS rules. Let $f'=\max\{ F|_{exks}\}$. We further have
\begin{eqnarray}
\tilde{F}|_{exks}\leq \lambda f'+L-1.
\label{new4}
\end{eqnarray}
Inequality (\ref{new2}) is valid for the value assignments obeying the KS rules, and inequality (\ref{new4}) is valid for the value assignments violating the KS rules. By combining Eqs. (\ref{new2}) and (\ref{new4}), we finally obtain the inequality
\begin{eqnarray}
\tilde{F} \leq \max\left\{\lambda f+L, \lambda f'+L-1\right\}.
\label{new5n}
\end{eqnarray}
Equation (\ref{new5n}) is always valid for all possible value assignments to the ray set, irrespective of whether the KS rules are obeyed or violated. Hence, we get an inequality without assuming the KS rules,
\begin{eqnarray}
\langle \tilde{F}\rangle \leq \max\left\{\lambda f+L, \lambda f'+L-1\right\}.
\label{new5}
\end{eqnarray}

Third, we discuss the quantum violation of the inequality. In quantum mechanics, each observable $P_i$ corresponds to an operator $\hat{P_i}=\ket{\varphi_i}\bra{\varphi_i}$. $\langle P_iP_j\rangle=0$ if $P_i$ and $P_j$ are orthogonal, and $\sum_{i=1}^n\langle P_i\rangle=\langle\sum_{i=1}^nP_i\rangle=1$ if $\{P_i, i=1,2,\cdots,n\}$ form an orthogonal basis. In quantum mechanics, the average value of $\tilde{F}$, expressed as Eq. (\ref{average}), reads
\begin{eqnarray}
\langle \tilde{F}\rangle_\psi =\lambda \langle F\rangle_\psi+L,
\label{averageq}
\end{eqnarray}
where the subscript $\psi$ indicates that the averages $\langle\tilde{F}\rangle_\psi$ and $\langle F\rangle_\psi$ are calculated by using the theory of quantum mechanics for state $\ket{\psi}$. By comparing Eqs. (\ref{new5}) and (\ref{averageq}), quantum violation occurs if
\begin{eqnarray}
\lambda \langle F\rangle_\psi+L>\max\left\{\lambda f+L, \lambda f'+L-1\right\}.
\label{new6}
\end{eqnarray}
The validity of Eq. (\ref{new6}) relies on the parameters $f$, $f'$, and $\lambda$. If both $f\geq f'$ and $\lambda>0$ or if both $f < f'$ and $0<\lambda\leq \frac{1}{f'-f}$, Eq. (\ref{new6}) can be guaranteed by
\begin{eqnarray}
\langle F\rangle_\psi>f,
\label{new7}
\end{eqnarray}
which is just the quantum violation of Eq. (\ref{eqF}). Therefore, there always exists $\lambda$ such that the quantum violation of Eq. (\ref{eqF}) can always guarantee the quantum violation of Eq. (\ref{new5}).

So far, we have proved that a KS inequality can always lead to a
noncontextuality inequality, and the latter is violated in quantum
mechanics if the former is violated. Note that the value assignments
of variables in a KS inequality are constrained by both the
assumption of noncontextuality and the KS rules while the
assignments in a noncontextuality inequality are constrained only by
the assumption of noncontextuality. The value assignments satisfying both the
assumption of noncontextuality and the KS rules are a subset of the value assignments satisfying only the
assumption of noncontextuality. If an inequality is valid for all the value assignments satisfying the assumption of noncontextuality, it must be valid for the subset of the value assignments, too. Therefore, a noncontextuality inequality itself can be regarded as a KS inequality. This completes the proof of the theorem that a noncontextuality inequality exists if and only if a KS inequality exists in a ray set.

It is interesting to see that the above proof has actually provided
an effective approach for constructing a noncontextuality inequality
with the aid of a KS inequality in any ray set. It is generally easier to
construct a KS inequality than a noncontextuality inequality in a
ray set.  Once a KS inequality is constructed, the theorem can
convert it to a noncontextuality inequality. By following the proof,
all the well-known KS inequalities proposed in the literature can be
converted to noncontextuality inequalities. For example, we consider
the first inequality proposed by Yu and Oh \cite{Yu.Oh2012}, with a
$13$-ray set in a three-dimensional quantum system. The
state-independent KS inequality [see Eq. (2) in Ref.
\cite{Yu.Oh2012}] reads $ \langle F\rangle_{ks} \le 1$, with
$F=\sum_{i=1}^{4}P_i$, where we have rewritten the expressions using
our notations. Here, $P_1, P_2,\dots,P_{13}$ are corresponding to
$h_0,h_2,h_2,h_3,z_2,y_2^+,y_2^-,z_3,y_3^+,y_3^-,z_1, y_1^+,y_1^-$
in Ref. \cite{Yu.Oh2012}, respectively. Starting from it, we can
immediately obtain a state-independent noncontextuality inequality
by following the process in the proof of the theorem. Indeed, by
comparing it with Eq. (\ref{eqF}), we have $f=1$, and by using the
definition $f'=\max\{F|_{exks}\}$, we have $f'=4$, which indicate
that $0<\lambda\leq\frac{1}{f'-f}= \frac{1}{3}$. The
state-independent noncontextuality inequality is given by Eq.
(\ref{new5}) as $\langle \tilde{F}\rangle \leq 4+\lambda$, where
$\tilde{F}$ is defined as Eq. (\ref{edefBj}) with $n=3$, $\mu=13$,
$L=4$ \footnote{There are four subsets of the rays that form a
three-dimensional orthogonal basis, $\mathcal{S}^1=\{P_5,P_6,P_7\}$,
$\mathcal{S}^2=\{P_8,P_9,P_{10}\}$,
$\mathcal{S}^3=\{P_{11},P_{12},P_{13}\}$, and
$\mathcal{S}^4=\{P_5,P_8,P_{11}\}$.}, $0<\lambda\leq \frac{1}{3}$,
and $F=\sum_{i=1}^{4}P_i$.

As shown in the above example, one may easily convert a KS inequality to a noncontextuality inequality, expressed as Eq. (\ref{new5}), by  calculating $f'$ and $\lambda$. However, the bound in expression (\ref{new5}) may not be tight, since it has been relaxed for the generality during the proof of the theorem. In a specific application, the tight bound, $\max\{\tilde{F}_{ks},\tilde{F}_{exks}\}$, can be obtained by detailedly calculating $\tilde{F}_{ks}$ and $\tilde{F}_{exks}$, and a tight inequality is given as $\langle \tilde{F}\rangle \leq \max\{\tilde{F}_{ks},\tilde{F}_{exks}\}$, instead of Eq. (\ref{new5}). We would like to illustrate this point by considering the KS inequality proposed by Klyachko \textit{et al.} in Ref. \cite{Klyachko.etal2008}, with a five-ray set in a three-dimensional quantum system. The state-dependent KS inequality [see Eq. (7) in Ref. \cite{Klyachko.etal2008}] reads $\langle F\rangle_{ks}\le 2$, with $F=\sum_{i=1}^5P_i$. Since the last term in Eq. (\ref{edefBj}) vanishes due to $L=0$ for this instance, Eq. (\ref{edefBj}) is reduced to $\tilde{F}=\lambda F-\sum_{i=1}^5P_iP_{i+1}$ , where the subscript $i$ is modulo $5$. In this case, we have $f=2$ , $f'=\max\{F|_{exks}\}=5$, and $0<\lambda\leq\frac{1}{f'-f}= \frac{1}{3}$. Substituting them into Eq. (\ref{new5}), we obtain $\langle \tilde{F}\rangle \le \max\{2\lambda,5\lambda-1\}$ with $0<\lambda\le \frac{1}{3}$, which is a state-dependent noncontextuality inequality but not a tight one. To find the tight bound, we calculate $\tilde{F}_{ks}$ and $\tilde{F}_{exks}$ detailedly, and have $\tilde{F}|_{ks}=0,\lambda,2\lambda$, $\tilde{F}|_{exks}=2\lambda-1,3\lambda-1,3\lambda-2,4\lambda-3,5\lambda-5$, and $\max\{\tilde{F}\}=\max\{\tilde{F}|_{ks},\tilde{F}|_{exks}\}=\max\{2\lambda, 3\lambda-1, 4\lambda-3, 5\lambda-5\}$. We then obtain the tight noncontextuality inequality, $\langle\tilde{F}\rangle\le \max\{2\lambda, 3\lambda-1, 4\lambda-3, 5\lambda-5\}$ with $0<\lambda\le 1$, where $0<\lambda\le 1$ is determined by requiring the quantum violation of the inequality. For $\lambda=1$, the inequality reads $\sum_{i=1}^5\langle P_i\rangle-\sum_{i=1}^5\langle P_iP_{i+1}\rangle\le 2$, which is exactly the KCBS inequality given in Ref. \cite{Klyachko.etal2008} if $P_i$ is replaced by $A_i$ with $P_i=\frac{1}{2}(1-A_i)$.

It should be noted that the above discussion is concerned essentially with a general non-KS set. If the ray set under consideration is a KS set, then the process of obtaining a noncontextuality inequality becomes much simpler. By definition, for a KS set, all possible value assignments to the set violate the KS rules. In this case, all value assignments obeying the KS rules certainly satisfy any given inequalities logically, because there are no value assignments obeying the KS rules at all, and therefore Eq. (\ref{eqF}) can be any form of inequality with a quantum violation. This implies that a state-independent noncontextuality inequality always exists for a KS set. However, it also implies that Eq. (\ref{eqF}) is useless to a KS set due to the arbitrariness of its form. Fortunately, the proof of the theorem provides a simple approach to obtain a state-independent noncontextuality inequality from a KS set. In fact, for a KS set, $\tilde F$ can be directly defined as
\begin{equation}
\tilde{F}=-\sum^\mu_{\substack{i>j=1,\\\hat{P_i}\hat{P_j}=0}}P_iP_j+
\sum_{\alpha=1}^L \left(\sum_{i=1}^n P_{k^\alpha_i}-2\sum_{i>j=1}^n P_{k^\alpha_i}P_{k^\alpha_j}\right),
  \label{edefBj2}
\end{equation}
and there is
\begin{eqnarray}
\tilde{F}|_{exks}\leq L-1.
\label{new42}
\end{eqnarray}
Noting that expression (\ref{new2}) does not need to be considered since all the value assignments to a KS set violate the KS rules, we then have
 \begin{eqnarray}
\langle \tilde{F}\rangle \leq L-1.
\label{new5b}
\end{eqnarray}
Equation (\ref{new5b}) is a state-independent noncontextuality inequality, since $\langle\tilde{F}\rangle_\psi=L~(>L-1)$ for any quantum state $\ket{\psi}$. Comparing with the previous method for constructing a noncontextuality inequality from a KS set, proposed by Badzi\c{a}g \textit{et al.} \cite{Badziag.etal2009}, our approach is much simpler. In the previous method, one needs to enlarge the given KS set to construct a state-independent noncontextuality inequality in general and correlations of $n$ compatible observables are involved, while only correlations of two compatible observables are involved in the present approach.

In passing, we point out that $P_i$ can be replaced by $A_i=1-2P_i$ in the proof of the theorem, and all of the above discussions work for $A_i$ although we have used $P_i$, instead of $A_i$, in this Rapid Communication for simplicity. By substituting $P_i=\frac{1}{2}(1-A_i)$ and $P_iP_j=\frac{1}{4}(1-A_i-A_j+A_iA_j)$, $\tilde F$ in Eqs.  (\ref{edefBj}) and (\ref{edefBj2})  can be written as a function of $A_i$, and the same procedure can be followed. However, the expressions as well as the calculations are much simpler by using $P_i$ than $A_i$. Besides, although our theoretical approach can help to determine the existence of a KS inequality in a ray set, the inequality obtained may not be optimal. The numerical approach proposed in Ref. \cite{Kleinmann.etal2012} may be used to optimize
it if needed.

In conclusion, we have established a theorem that a noncontextuality
inequality exists if and only if a KS inequality exists in a ray
set. The theorem not only solves the coexistence of the two types of
inequalities in any ray set, but also provides an effective approach
both for constructing noncontextuality inequalities in a KS set and
for converting a KS inequality to a noncontextuality inequality in
any ray set. A noncontextuality inequality is immediately obtained
when the approach is applied to a KS set, and all the known KS inequalities
can be converted to noncontextuality inequalities when it is
applied to any ray set.

\begin{acknowledgments}
We thank Murray Batchelor for useful discussions. This work was supported by NSF China by Grant No.11175105 and the Taishan Scholarship Project of Shandong Province. D.M.T acknowledges support from the National Basic Research Program of China by Grant No. 2009CB929400.

\end{acknowledgments}

%\bibliography{contextuality}


\end{document}